\documentclass[twocolumn,journal]{IEEEtran}

\usepackage{graphicx}
\usepackage{amsmath}
\usepackage{amssymb}

\usepackage{amsthm}
\usepackage{cite}
\usepackage{float}
\usepackage{balance}

\begin{document}

\title{On the Coexistence of a Primary User with an Energy Harvesting Secondary User: A Case of Cognitive Cooperation}
\author{Ahmed El Shafie$^\dagger$, Tamer Khattab$^*$, Amr El-Keyi$^\dagger$, Mohamed Nafie$^\dagger$\\
\small \begin{tabular}{c}
$^\dagger$Wireless Intelligent Networks Center (WINC), Nile University, Giza, Egypt. \\
$^*$Electrical Engineering, Qatar University, Doha, Qatar. \\
\end{tabular}
\thanks{Mohamed Nafie is also affiliated with the Electronics and Communications Department, Cairo University.}
}
\date{}
\maketitle
\thispagestyle{empty}
\pagestyle{empty}
\begin{abstract}
In this paper, we consider a cognitive scenario where an energy harvesting secondary user (SU) shares the spectrum with a primary user (PU). The secondary source helps the primary source in delivering its undelivered packets during periods of silence of the primary source. The primary source has a queue for storing its data packets, whereas the secondary source has two data queues; a queue for storing its own packets and the other for storing the fraction of the undelivered primary packets accepted for relaying. The secondary source is assumed to be a battery-based node which harvests energy packets from the environment. In addition to its data queues, the SU has an energy queue to store the harvested energy packets. The secondary energy packets are used for primary packets decoding and data packets transmission. More specifically, if the secondary energy queue is empty, the secondary source can neither help the primary source nor transmit a packet from the data queues. The energy queue is modeled as a discrete time queue with Markov arrival and service processes. Due to the interaction of the queues, we provide inner and outer bounds on the stability region of the proposed system. We investigate the impact of the energy arrival rate on the stability region. Numerical results show the significant gain of cooperation.
\end{abstract}
\section{Introduction}
 The electromagnetic radio spectrum is a scarce resource \cite{haykin2005cognitive}. Regulatory bodies have announced that the electromagnetic spectrum is left unused for most of the time and large portions  of certain licensed frequency bands remain
idle. Recently, researchers have proposed the idea of cognitive radio, whereby the secondary users (SUs) can use the licensed primary resources whenever these resources are unutilized by the primary users (PUs).

Designing an energy-efficient scheme for wireless sensor networks is desirable due to its practical applications \cite{ref1,ref2,ref3}. This is because sensor nodes are generally battery-operated and therefore energy consumption is very important. In many practical applications, the SU is an energy constrained device which is equipped by a rechargeable battery. The secondary operation, which involves spectrum sensing, packets decoding, and channel accessing, is accompanied by energy consumption. Consequently, for designing an energy efficient system, the SU must optimize its decisions to efficiently invest the available energy in its battery.

Energy harvesting technology is an emerging technology that has been recognized as
a promising effective solution to optimally increase the lifetime
of wireless energy constrained networks by eliminating the
cost for hard-wiring or replacing batteries of rechargeable mobile devices \cite{c0}. It enables wireless nodes to collect (harvest) energy
from the surrounding environment \cite{c1}. For a comprehensive overview of the different energy harvesting technologies, the reader is referred to \cite{survey} and the references therein.

Several articles have discussed optimal energy management strategies such as \cite{hoang2009opportunistic,sharma2010optimal,ho2010optimal,yang2010transmission,yang2010optimal,tutuncuoglu2010optimum}. The authors of \cite{hoang2009opportunistic} investigate an energy constrained
cognitive terminal without explicitly involving an
energy queue. Throughput and mean delay optimal energy management policies for single-node communication were identified in \cite{sharma2010optimal}. The energy allocation over a finite horizon with the objective of maximizing the throughput was investigated in \cite{ho2010optimal}.
Minimizing the transmission completion time of an
energy harvesting system was considered in \cite{yang2010transmission} and \cite{yang2010optimal}. The optimal solution is obtained
using a geometric framework.
 The authors in \cite{tutuncuoglu2010optimum} considered energy harvesting transmitters with
batteries of finite energy storage capacity and
the problem of throughput maximization by a deadline is
solved for a static channel.

  The authors of \cite{pappas2012optimal} studied the stability region, which describes the theoretical limit on data rates of data queues that can be pushed into the system while maintaining all the data queues stable, for a cognitive radio network consisting of one rechargeable PU and one SU plugged to a reliable power source under the assumption of a decoupled ${\rm M/D/1}$\footnote{The notation of discrete-time ${\rm M/D/1}$ queue is used to describe a queueing
system with Bernoulli arrival process and deterministic service process.} energy queue with unity service rate. This assumption implies one energy packet expenditure each time slot regardless of state of all other queues in the network. The impact of cooperation on the stable throughput of a source in a wireless three-node network topology (source-relay-destination) was investigated in \cite{krikidis2012stability}. The nodes were energy harvesters with bursty data traffic and without channel state information (CSI) at the transmitters. The authors in \cite{Sult1210:Optimal} characterized the maximum stable throughput for a single secondary terminal sharing the channel with a single primary terminal. The SU randomly senses and accesses the channel to achieve the maximum secondary throughput. Sultan \cite{Sultan} investigated the optimal sensing
and access policies for an energy harvesting SU based on the Markov decision process (MDP). In \cite{gc2013}, the authors investigated the optimal sensing duration that an energy harvesting SU selects in each time slot to maximize its stable throughput. The primary and secondary terminals are assumed to be energy harvesters.

Cooperation between the primary and the secondary systems has gained considerable attention recently \cite{ref4,ref5,ref6,ref7,simeone,khattab,erph}. Under cooperation, the SU aids the PU to increase the communication reliability of its transmissions and overcome the channel fading effects. The cooperation is useful for the secondary source to increase the spectrum availability for its own transmissions.

In \cite{simeone}, the authors study a cognitive network with one primary and one secondary transmitter. The secondary source adjusts its power such that the secondary queue mean service rate is maximized while preserving the stability of both the primary and secondary queues. Elsaadany {\it et al.} \cite{khattab} considered a network with one PU and one SU capable of relaying. The secondary delay is minimized under constraints that ensure the stability of the primary, the secondary, and the relaying queues and for a given predefined threshold on the relaying power
budget. The authors of \cite{erph} characterized the stability region in a two user cognitive network where the nodes are supplied with multipacket reception capabilities.

In this work, we study the impact of having a cooperative energy harvesting SU with the PU on the stability region. The analysis is carried out from network layer perspective. We make the following contributions in this paper.
\begin{itemize}
\item We investigate the stability region of an energy harvesting cooperative SU. We provide inner and outer bounds on the stability region.
    \item The proposed system is studied with two energy queue models. In the first model, we propose to model the energy queue as a discrete-time queue with certain arrival rate and certain service rate. The service process of the energy queue depends on the state of the data queues and the secondary operation (channel accessing and primary packet decoding). In the second model, we make use of the decoupled ${\rm M/D/1}$ queue with unity service rate which was assumed in, e.g., \cite{pappas2012optimal,Sult1210:Optimal}, and the references therein. The ${\rm M/D/1}$ with unity service rate model assumes a packet consumption per slot regardless of the other queues state and the secondary operation and therefore it is a non-realistic assumption. This is because the energy consumption could happen every slot even if the SU is silent (in terms of decoding a primary packet or accessing a sensed free time slot with a data packet).
          \item Numerically, we study the impact of the energy arrival rate on the stability region of the network.
\end{itemize}

The rest of the paper is organized as follows. In the next section,
we describe the system model considered in this paper. The stability region of the proposed system is investigated in Section \ref{stability}.
In Section \ref{numerical}, we provide some numerical results for the optimization problems considered in this paper. The conclusions are drawn in Section \ref{conc}.
\section{SYSTEM MODEL} \label{model}
\begin{figure}[t]
    \centering
\normalcolor
  \includegraphics[width=1\columnwidth]{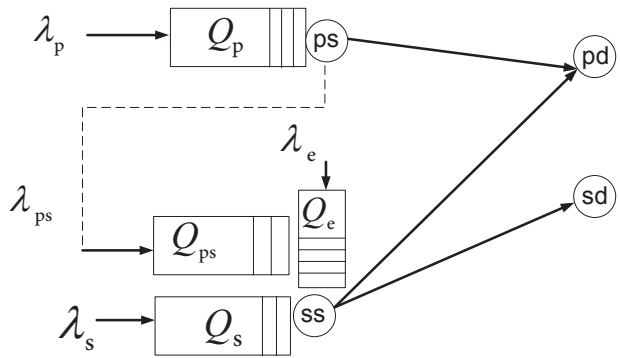}
  \caption{Primary and secondary links and queues. We denote the primary source as `${\rm ps}$', the primary destination as `${\rm pd}$', the secondary source as `${\rm ss}$', and the secondary destination as `${\rm sd}$'. In the figure, $\lambda_{\rm ps}$ denotes the average arrival rate to the relaying queue.}\label{sysmod}
\end{figure}

We consider a cognitive scenario composed of one PU and one energy harvesting SU as shown in Fig. \ref{sysmod}. The proposed setting shown in Fig. \ref{sysmod} can be seen as a subsystem in a larger network in which the primary and secondary pairs are sharing orthogonal channels. The analysis provided in this paper focuses on the pair using the same channel. The SU is utilizing the
spectrum resource whenever it has energy for transmission and the primary source is sensed to be inactive (its transmission queue is empty). The secondary source senses the channel every time slot to detect possible activity of the primary source. In this work, the sensing process is assumed to be perfect \cite{erph,khattab}. That is, the probabilities of misdetction and false-alarm are negligible due to the use of large sensing duration $\tau$. The secondary source will be able to transmit a packet each time slot during the idle sessions of the primary source. An energy packet is consumed either in data packet transmission or in primary data packet decoding and in feedback messages broadcasting.

We adopt a flat fading channel model and assume that the channel gains remain constant over the duration of the time slot. We do not assume the availability of the CSI at the transmitting terminals. Let us denote the primary source as `${\rm ps}$', the primary destination as `${\rm pd}$', the secondary source as `${\rm ss}$', and the secondary destination as `${\rm sd}$'. Let $h^t_{j,k}$ denote the channel coefficient between node $j$ and node $k$ in time slot $t$, where $j,k \in \{\rm ss,sd,ps,pd\}$. We assume an erasure channel, where a packet is successfully decoded by the destination if the received signal-to-noise-ratio (SNR) at the receiver exceeds the correct decoding threshold, or equivalently, if the rate of the transmitted packet is lower than channel capacity \cite{sadek}. Since we adopt a time slotted system with a fixed packet transmission rate, the correct packet reception will be captured in terms of outage events and outage probability \cite{sadek}. The received signal at node $k$ due to a packet transmission from node $j$ in time slot $t$ can be modeled as
\begin{equation}\label{modelll}
    y_{j,k}^t=h_{j,k}^t \mathcal{X}_j^t+z_{k}^t
\end{equation}
\noindent where $\mathcal{X}^t_j$ denotes the transmitted packet with average power $\mathcal{P}_{j,k}$ Watts at time $t$, and $z^t_{k}$ is the additive white Gaussian noise (AWGN) at node $k$.
Based on the system model, the outage probability (complement of successful transmission probability) between node $j$ and node $k$ can be calculated as follows \cite{sadek,krikidis2012stability}:
\begin{equation}
\label{outage}
\begin{split}
{\rm Pr}\{O_{j,k}\}\!=\!P_{j,k}&\!=\!{\rm Pr}\bigg\{r_j\!>\!C_{j,k}\!=\!W\log_2\bigg(1\!+\!|h^t_{j,k}|^2\frac{\mathcal{P}_{j,k}}{\mathcal{N}_\circ}\bigg)\bigg\}\\&={\rm Pr}\bigg\{|h^t_{j,k}|^2\frac{\mathcal{P}_{j,k}}{\mathcal{N}_\circ}< 2^{\frac{r_j}{W}}-1\bigg\}
\\& =\!\!1\!-\!\exp\biggr(\!-\!\frac{2^{\frac{r_j}{W}}\!-\!1}{\gamma_{j,k}}\!\biggr)
\end{split}
\end{equation}
\noindent where $O_{j,k}$ is the event that the channel between node $j$ and node $k$ is in outage, $r_j$ is the transmission rate of terminal $j$, $W$ is the channel bandwidth, $C_{j,k}$ is the capacity of the link between node $j$ and node $k$, $\mathcal{N}_\circ$ is the AWGN power in Watts, $\sigma^2_{j,k}$ is the mean value of the channel gain $|h^t_{j,k}|^2$, and $\gamma_{j,k}= \sigma^2_{j,k}\mathcal{P}_{j,k}/\mathcal{N}_\circ$ is the received SNR at node $k$ when node $j$ transmits a packet to node $k$.
The probability of correct packet reception between node $j$ and node $k$ is $\overline{P}_{j,k}~=~1~-~P_{j,k}$. The SU expends one energy packet for data transmission. Assume that the energy contained in one energy packet is $\mathbb{E}$ energy units. Let $T$ denote the time slot length. Since the secondary transmission time is $T-\tau$, the secondary transmit power is $\mathcal{P}_{\rm ss,pd}=\mathcal{P}_{\rm ss,sd}=\mathbb{E}/(T-\tau)$.

We assume that the primary source has an infinite length buffer to store the arrived packets to the primary queue, denoted by $Q_{\rm p}$. The secondary source has three buffers. Specifically, a buffer to keep its own packets, denoted by $Q_{\rm s}$, and another buffer to keep the relaying packets, denoted by $Q_{\rm ps}$, and the third buffer maintains the energy harvested from the environment, denoted by $Q_{\rm e}$. The buffers are assumed to be of
infinite length which is a reasonable assumption if the packet length is very small relative to buffers' size \cite{krikidis2012stability}. We consider time-slotted transmissions where all packets contain the same number of bits and one slot is sufficient for one packet transmission. More specifically, a packet of length $b$ bits is transmitted in $T$ seconds. Accordingly, the transmission rate of the primary source is $r_{\rm p}=b/T$, whereas the transmission rate of the secondary source is $r_{\rm s}=b/(T-\tau)=r_{\rm p}/(1-\tau/T)$, where $\tau$ is the time spent in channel sensing and is assumed to be a fraction of $T$.\footnote{The sensing duration $\tau$ should be long enough to ensure robust channel sensing outcomes. However, the delay in secondary transmission due to channel sensing raises the channel outages and hence the queues rates degrade (for proof, see \cite{Sult1210:Optimal}).} The arrivals to the primary queue and the secondary data an energy queues are assumed to be independent Bernoulli (Markov) random variables \cite{krikidis2012stability} with mean arrival rates $\lambda_{\rm p}\in [0,1]$ packets/slot, $\lambda_{\rm s} \in [0,1]$ packets/slot, and $\lambda_{\rm e}\in [0,1]$ energy packets/slot for the primary queue, secondary queue, and secondary energy queue, respectively. The arrivals are assumed to be independent from queue to another and from slot to slot \cite{sadek,krikidis2012stability}.

The feedback messages are assumed to be instantaneous and error-free\footnote{This is because acknowledgement (ACK) and negative-acknowledgement (NACK) packets are very short augmented by the use of low rate and strong codes \cite{sadek,krikidis2012stability}.}, and all nodes in the system can
hear them.
We summarize medium access control (MAC) operation as follows.
\begin{itemize}
                                                                 \item If the primary queue is nonempty, the primary source transmits the packet at the head of its queue at the beginning of each time slot.
                                                                     \item The secondary source senses the primary channel each time slot to detect possible activity of the primary source.
                                                                             \item If the primary source is active, the secondary source attempts to decode a fraction $f$ of the primary packets if and only if it has energy packets.
                                                                 \item If a packet is received successfully by either the primary destination
or the secondary source, a feedback message is sent at the end of the time slot by the receiving node to inform the primary source regarding the packet decoding status. The packet is then removed from the primary source's queue and secondary source's relaying queue if the receiving node was the primary destination.
    \item If the primary destination could not receive the primary packet correctly and the secondary source correctly decoded and accepted the packet, then the secondary source will send back an ACK message at the end of the slot and the packet will be dropped from the primary queue.
                                                                 \item If a packet is not received successfully by both the primary destination and
the secondary source, each of them sends back a NACK message\footnote{We assume that the ACKs/NACKs sent by the secondary source and the primary destination are separated in time or frequency.}, and the primary source retransmits this packet in the next time slot.
                                                                 \item If the channel is sensed to be free of primary activity, which occurs when the PU's queue is empty, the secondary source, if it has energy in its energy queue, either retransmits the packet at the head of $Q_{\rm ps}$ or transmits the packet at the head of $Q_{\rm s}$.
                                                               \end{itemize}

\begin{table}
\renewcommand{\arraystretch}{1}
\begin{center}
\begin{tabular}{ |c |l| }
    \hline\hline
        ${\rm ss}$ & {\footnotesize Secondary source} \\[5pt]\hline
        ${\rm ps}$ & {\footnotesize Primary source} \\[5pt]\hline
        ${\rm sd}$ & {\footnotesize Secondary destination} \\[5pt]\hline
        ${\rm pd}$ & {\footnotesize Primary destination} \\[5pt]\hline

    $\tau$ & {\footnotesize Sensing duration} \\[5pt]\hline
  $T$ & {\footnotesize Slot duration} \\[5pt]\hline
    $W$ & {\footnotesize Channel bandwidth} \\[5pt]\hline
    $\mathbb{E}$ & {\footnotesize Amount of energy per energy packet} \\[5pt]\hline

      $h_{j,k}$ & {\footnotesize Channel coefficient between node $j$ and node $k$} \\[5pt]\hline
  $r_j$ & {\footnotesize Transmission rate of node $j$} \\[5pt]\hline
        $\mathcal{C}_{j,k}$ & {\footnotesize Channel capacity between node $j$ and node $k$} \\[5pt]\hline
  $\mathcal{N}_\circ$ & {\footnotesize Noise power in Watts} \\[5pt]\hline
      $\gamma_{j,k}$ & {\footnotesize Received SNR at receiver $k$} \\[5pt]\hline
      $\mathcal{P}_{j,k}$ & {\footnotesize Transmit power of node $j$ while communicating with node $k$} \\[5pt]\hline

  $\beta$ & {\footnotesize Probability of choosing $Q_{\rm s}$ packets} \\& { for transmission when $Q_{\rm ps}\ne0$} \\[5pt]\hline
   $1-\beta$ & {\footnotesize Probability of choosing $Q_{\rm ps}$ packets} \\& {for transmission when $Q_{\rm s}\ne0$} \\[5pt]\hline
    $Q_{\rm p}$ & {\footnotesize Primary queue} \\[5pt]\hline
  $Q_{\rm s}$ & {\footnotesize Secondary traffic queue} \\[5pt]\hline
  $Q_{\rm ps}$ & {\footnotesize Relaying queue} \\[5pt]\hline
     $Q_{\rm e}$ & {\footnotesize Energy queue} \\[5pt]\hline
         $\lambda_\ell$ & {\footnotesize Mean arrival rate of $Q_\ell$} \\[5pt]\hline
         $\mu_\ell$ & {\footnotesize Mean service rate of $Q_\ell$} \\[5pt]\hline

    $f$ & {\footnotesize The probability of admitting a primary packet} \\[5pt]\hline
  $\overline{P}_{\rm ps,pd}$ & {\footnotesize Probability of successful primary transmission} \\& { to the primary destination}\\[5pt]\hline
   $\overline{P}_{\rm ps,ss}$ & {\footnotesize Probability of successful primary transmission} \\& { to the secondary source}\\[5pt]\hline
 $\overline{P}_{\rm ss,sd}$ & {\footnotesize Probability of successful secondary transmission} \\& { to the secondary destination}\\[5pt]\hline
   $\overline{P}_{\rm ss,pd}$ & {\footnotesize Probability of successful secondary transmission} \\& { to the primary destination}\\[5pt]\hline
\end{tabular}
\caption{List of Symbols.}
\label{tablett}
\end{center}
\end{table}
\section{Stability analysis of the system}\label{stability}
 The queue sizes of $Q_\ell$, $\ell\in\{\rm p,s,ps,e\}$, evolves according to the following equation:
\begin{equation}\label{queue}
    Q_\ell^{t+1}=\bigr(Q_\ell^t-\mathcal{U}_\ell^t\bigr)^{+}+\mathcal{A}^t_\ell
\end{equation}
\noindent where $\mathcal{U}_\ell^t$ is the number of departures in time slot $t$ and $\mathcal{A}^t_\ell$ denotes the number of arrivals in time slot $t$ and is a stationary process by assumption with finite mean $\mathcal{E}\{\mathcal{A}^t_\ell\}=\lambda_\ell$. The function $(.)^{+}$ is defined as $(\theta)^{+}=\max\{\theta,0\}$, where $\max\{\cdot\}$ returns the maximum of the values provided in the argument. We assume that
departures occur before arrivals, and the queue size is measured at the beginning of the time slot \cite{sadek}.

\emph{Definition:} A data queue $Q_\ell$, where $\ell \in\{\rm p,s,ps\}$, is stable if
\begin{equation}\label{stabilityeqn}
    \lim_{\mathcal{K} \rightarrow \infty} \lim_{t \rightarrow \infty  }{\rm Pr}\bigg\{Q_\ell^t<\mathcal{K}\bigg\}=1
\end{equation}
\noindent All the data queues in the network should be stable for proper operation of the system. We can apply Loynes' theorem to check for stability conditions \cite{sadek}, if the arrival and service processes are strictly stationary. Loynes' theorem states that if the arrival process and the service process of a queue are strictly stationary processes, and the mean service rate is greater than the mean arrival rate of the queue, then the queue is stable. If the mean service rate is lower than the mean arrival rate, then the queue is unstable. Mathematically, the data queue $Q_\ell$ is stable if $\lambda_\ell < \mu_\ell$, where $\lambda_\ell$ and $\mu_\ell$ are the mean arrival and service rates of $Q_\ell$, respectively.

Now, we move our attention to the queues arrival and service rates. The meaning of the relevant symbols are provided in Table \ref{tablett}. The secondary source exploits the sensed free time slots to transmit a packet from the relaying queue with probability $\overline{\beta}$ when $Q_{\rm s}\ne0$, and with probability $1$ when $Q_{\rm s}=0$. Or it transmits a packet from its own traffic with probability $\beta=1-\overline{\beta}$ when $Q_{\rm ps}\ne0$, and with probability $1$ when $Q_{\rm ps}=0$. Based on the system model, the service rate of the queues are given as follows. When the secondary energy queue is nonempty, a packet from the primary queue, $Q_{\rm p}$, is served in either one of the following events. If its channel is not in outage (which occurs with probability $\overline{P}_{\rm ps,pd}$); or if its channel is in outage, the channel between the primary source and the secondary source is not in outage (which occurs with probability $\overline{P}_{\rm ps,ss}$), and the secondary source decides to accept the packet (which occurs with probability $f$). Therefore, the mean service rate of the primary queue is given by
\begin{equation}
 \mu_{\rm p} \!=\! \overline{P}_{\rm ps,pd}\!+\!f P_{\rm ps,pd}\overline{P}_{\rm ps,ss} {\rm Pr}\bigg\{Q_{\rm e}\!\ne\!0\bigg\}
 \end{equation}

Accordingly, the mean arrival rate to the relaying queue, $Q_{\rm ps}$, is given by
 \begin{equation}
 \label{77}
   \lambda_{\rm ps}=f P_{\rm ps,pd}\overline{P}_{\rm ps,ss}{\rm Pr}\bigg\{Q_{\rm p}\ne0,Q_{\rm e}\ne0\bigg\}
   \end{equation}

   A packet from the relaying queue, $Q_{\rm ps}$, is served if the primary queue is empty, the secondary source has energy in its energy queue, $Q_{\rm s}=0$ or $Q_{\rm s}\ne0$ and the secondary source decides to transmit from the relaying queue (which occurs with probability $\overline{\beta}$), and the channel between the secondary source and the primary destination is not in outage. Thus, the mean service rate of the relaying queue is given by
   \begin{equation}
   \begin{split}
   \label{88}
  \mu_{\rm ps}&=  \overline{P}_{\rm ss,pd}\biggr({\rm Pr}\bigg\{Q_{\rm p}=0,Q_{\rm s}=0,Q_{\rm e}\ne0\bigg\}\\& \,\,\,\,\,\,\,\,\,\,\,\,\,\,\,\ +\overline{\beta}{\rm Pr}\bigg\{Q_{\rm p}=0,Q_{\rm s}\ne0,Q_{\rm e}\ne0\bigg\}\biggr)
  \end{split}
  \end{equation}

  Now we move our attention to the secondary packets queue, $Q_{\rm s}$. A packet at the head of this queue is served in either one of the following events. If the primary queue is empty, the secondary source has energy in its energy queue, $Q_{\rm ps}=0$ or $Q_{\rm ps}\ne0$ and the secondary source chooses its own data queue for transmission (which occurs with probability $\beta$), and the channel between the secondary source and the secondary destination is not in outage. Therefore, the mean service rate of $Q_{\rm s}$ is given by
  \begin{equation}
  \label{99}
  \begin{split}
  \mu_{\rm s} &= \overline{P}_{\rm ss,sd}\biggr({\rm Pr}\bigg\{Q_{\rm p}=0,Q_{\rm ps}=0,Q_{\rm e}\ne0\bigg\}\\& \,\,\,\,\,\,\,\,\,\,\,\,\,\,\,\,\,\ +\beta{\rm Pr}\bigg\{Q_{\rm p}=0,Q_{\rm ps}\ne0,Q_{\rm e}\ne0\bigg\}\biggr)
  \end{split}
  \end{equation}

  A packet from the energy queue, $Q_{\rm e}$, is consumed in either one of the following events. If the PU is active and the SU decides to decode its packet (which occurs with probability $f$); or if the PU is inactive and the SU has a packet at any of its data queues. Mathematically,
   \begin{equation}
    \begin{split}
  \label{mueexp}
  \mu_{\rm e} &= f {\rm Pr}\bigg\{Q_{\rm p}\ne0\bigg\}\\& +\biggr({\rm Pr}\bigg\{Q_{\rm p}\!=\!0,Q_{\rm s}\!=\!0,Q_{\rm ps}\!\ne\!0\bigg\}\!\\& \,\,\,\ +\!{\rm Pr}\bigg\{Q_{\rm p}\!=\!0,Q_{\rm s}\!\ne\!0,Q_{\rm ps}\!=\!0\bigg\}\!\\& \,\,\,\ +\!{\rm Pr}\bigg\{Q_{\rm p}\!=\!0,Q_{\rm ps}\!\ne\!0,Q_{\rm ps}\!\ne\!0\bigg\}\biggr)
  \end{split}
  \end{equation}

  The queues are interacting with each other. More specifically, the service rates of the queues depend on each other and therefore the rates of the individual departure processes cannot be computed directly. Analyzing the stability
of interacting queues is a difficult problem that has been
addressed for ALOHA systems initially in \cite{tsy}, and later in
\cite{rao1988stability,szpankowski1994stability,luo1999stability,naware2005stability}. Characterizing the stable
throughput region for more than three interacting queues is still an open problem \cite{sadek}. To bypass this difficulty and render the characterization of the stability region tractable, we construct {\bf hypothetical} systems via bounding the queues' rates. In the new systems, the queues are decoupled.

Before proceeding, we emphasize the following. {\it Any point (rate pair $(\lambda_{\rm p},\lambda_{\rm s})$) beneath the outer bound is either stable or unstable, but all points above the outer bound are unstable. Moreover, all points beneath the the inner bound are stable, but any point above the inner bound is either stable or unstable.}
\subsection{Lower and upper bounds on the service and arrival rates}
We append superscript `$i$' and `$o$' to the notations of the mean arrival and service rates to denote the inner and the outer bounds, respectively. By definition, the mean service rate of queue $Q_\zeta$ should satisfy the inequality $\mu_\zeta^{\left(i\right)}\le\mu_\zeta\le \mu_\zeta^{\left(o\right)}$. In this subsection, we attempt to obtain upper and lower bounds on the queues' mean rates (arrival and departure) by constructing {\bf hypothetical} systems in which the queues' rates become bounded and decoupled. For the decoupled queues, the joint probabilities of queues state are simply the product of the marginal probabilities. The probability of a stable data queue $Q$ with mean arrival rate $\lambda$ and mean service rate $\mu^{\left(m\right)}\ge \lambda$, $m=\{i,o\}$, being nonempty is $\lambda/\mu^{\left(m\right)}$ \cite{sadek,krikidis2012stability,simeone,khattab,erph}. Therefore, the probability of $Q_\zeta$, $\zeta\in\{\rm p,s\}$, being empty is bounded by
  \begin{equation}
    \begin{split}
  \label{99}
  \frac{\lambda_\zeta}{\mu_\zeta^{\left(o\right)}}\le {\rm Pr}\bigg\{Q_\zeta\ne0\bigg\}\le \frac{\lambda_\zeta}{\mu_\zeta^{\left(i \right)}}
  \end{split}
  \end{equation}

  The mean service rate of the primary queue can be upper bounded as follows:
\begin{equation}
\begin{split}
 \mu_{\rm p} &= \overline{P}_{\rm ps,pd}+f P_{\rm ps,pd}\overline{P}_{\rm ps,ss} {\rm Pr}\bigg\{Q_{\rm e}\ne0\bigg\}\\&  \le \overline{P}_{\rm ps,pd}+f P_{\rm ps,pd}\overline{P}_{\rm ps,ss}=\mu_{\rm p}^{\left(o \right)}
 \end{split}
 \end{equation}

  The mean service rate of the energy queue $Q_{\rm e}$ is lower bounded as follows:
  \begin{equation}
    \begin{split}
  \mu_{\rm e} &= f {\rm Pr}\bigg\{Q_{\rm p}\ne0\bigg\}\\& +\biggr({\rm Pr}\bigg\{Q_{\rm p}\!=\!0,Q_{\rm s}\!=\!0,Q_{\rm ps}\!\ne\!0\bigg\}\!\\&  +\!{\rm Pr}\bigg\{Q_{\rm p}\!=\!0,Q_{\rm s}\!\ne\!0,Q_{\rm ps}\!=\!0\bigg\}\! \\ &  +\!{\rm Pr}\bigg\{Q_{\rm p}\!=\!0,Q_{\rm ps}\!\ne\!0,Q_{\rm ps}\!\ne\!0\bigg\}\biggr)\\& \ge f {\rm Pr}\bigg\{Q_{\rm p}\ne0\bigg\}\ge f \frac{\lambda_{\rm p}}{\mu_{\rm p}^{\left(o \right)}}= \mu_{\rm e}^{\left(i \right)}
  \end{split}
  \end{equation}
  with ${\lambda_{\rm p}}\le{\mu^{\left(o \right)}_{\rm p}}$. For the energy queue, if $\lambda_{\rm e}\le \mu^{\left(m \right)}_{\rm e}$, $m=\{i,o\}$, the probability of having the energy queue nonempty is given by $\frac{\lambda_{\rm e}}{\mu^{\left(m \right)}_{\rm e}}$. Whereas, if  $\lambda_{\rm e}> \mu^{\left(m \right)}_{\rm e}$, the energy queue becomes nonempty all the time. Combining both cases, the probability of having the energy queue nonempty is upper and lower bounded as
    \begin{equation}
    \begin{split}
\!\min\bigg\{\frac{\lambda_{\rm e}}{\mu_{\rm e}^{\left(o \right)}},1\bigg\} \le    {\rm Pr}\bigg\{Q_{\rm e}\ne0\bigg\} \le \min\bigg\{\frac{\lambda_{\rm e}}{\mu_{\rm e}^{\left(i \right)}},1\bigg\}
  \end{split}
  \end{equation}
  where $\min\{\cdot\}$ returns the minimum of the values in the argument. Using the following fact,
     \begin{equation}
   \begin{split}
   \label{88xx}
&{\rm Pr}\bigg\{Q_{\rm p}\!=\!0,Q_{\rm s}\!=\!0,Q_{\rm e}\!\ne\!0\bigg\}\!+\!\overline{\beta}{\rm Pr}\bigg\{Q_{\rm p}\!=\!0,Q_{\rm s}\!\ne\!0,Q_{\rm e}\!\ne\!0\bigg\}\\\! & \le\!  {\rm Pr}\bigg\{Q_{\rm p}\!=\!0,Q_{\rm s}\!=\!0,Q_{\rm e}\!\ne\!0\bigg\}\!+\!{\rm Pr}\bigg\{Q_{\rm p}\!=\!0,Q_{\rm s}\!\ne\!0,Q_{\rm e}\!\ne\!0\bigg\}\\& = {\rm Pr}\bigg\{Q_{\rm p}=0,Q_{\rm e}\ne0\bigg\}
  \end{split}
  \end{equation}
  The relaying queue mean service rate can be upper bounded as follows:
   \begin{equation}
   \begin{split}
   \label{88}
  \mu_{\rm ps}&=  \overline{P}_{\rm ss,pd}\biggr({\rm Pr}\bigg\{Q_{\rm p}=0,Q_{\rm s}=0,Q_{\rm e}\ne0\bigg\}\\& \,\,\,\,\,\,\,\,\,\,\,\,\,\,\,\,\,\,\,\,\,\ +\overline{\beta}{\rm Pr}\bigg\{Q_{\rm p}=0,Q_{\rm s}\ne0,Q_{\rm e}\ne0\bigg\}\biggr)\\& \le \overline{P}_{\rm ss,pd}{\rm Pr}\bigg\{Q_{\rm p}=0,Q_{\rm e}\ne0\bigg\}\!\\&\le \overline{P}_{\rm ss,pd}(1-\frac{\lambda_{\rm p}}{\mu^{\left(o \right)}_{\rm p}})\min\bigg\{\frac{\lambda_{\rm e}}{\mu_{\rm e}^{\left(i \right)}},1\bigg\}=\! \mu_{\rm ps}^{\left(o \right)}
  \end{split}
  \end{equation}
with ${\lambda_{\rm p}}\le{\mu^{\left(o \right)}_{\rm p}}$.

  The relaying queue mean arrival rate can be lower bounded as follows:
   \begin{equation}
   \begin{split}
   \lambda_{\rm ps}&=f P_{\rm ps,pd}\overline{P}_{\rm ps,ss}{\rm Pr}\bigg\{Q_{\rm p}\ne0,Q_{\rm e}\ne0\bigg\}\\& \ge f  P_{\rm ps,pd}\overline{P}_{\rm ps,ss}\min\bigg\{\frac{\lambda_{\rm e}}{\mu_{\rm e}^{\left(o \right)}},1\bigg\}\frac{\lambda_{\rm p}}{\mu^{\left(o \right)}_{\rm p}}\!=\!  \lambda_{\rm ps}^{\left(i \right)}
   \end{split}
   \end{equation}
with ${\lambda_{\rm p}}\le{\mu^{\left(o \right)}_{\rm p}}$.

   The secondary own data queue mean service rate is upper bounded as follows:
  \begin{equation}
  \begin{split}
  \label{99}
  \mu_{\rm s} &= \overline{P}_{\rm ss,sd}\biggr({\rm Pr}\bigg\{Q_{\rm p}=0,Q_{\rm ps}=0,Q_{\rm e}\ne0\bigg\}\\& \,\,\,\,\,\,\,\,\,\,\,\,\,\,\,\,\,\,\,\,\,\ +\beta{\rm Pr}\bigg\{Q_{\rm p}=0,Q_{\rm ps}\ne0,Q_{\rm e}\ne0\bigg\}\biggr)\\& \le \overline{P}_{\rm ss,sd}(1-\frac{\lambda_{\rm p}}{\mu^{\left(o \right)}_{\rm p}}) \min\bigg\{\frac{\lambda_{\rm e}}{\mu_{\rm e}^{\left(i \right)}},1\bigg\}=\!\mu_{\rm s}^{\left(o \right)}
  \end{split}
  \end{equation}
with ${\lambda_{\rm p}}\le{\mu^{\left(o \right)}_{\rm p}}$.

The primary mean service rate is lower bounded as follows:
\begin{equation}
\begin{split}
 \mu_{\rm p} \!&=\! \overline{P}_{\rm ps,pd}\!+\!f P_{\rm ps,pd}\overline{P}_{\rm ps,ss} {\rm Pr}\bigg\{Q_{\rm e}\!\ne\!0\bigg\}\!\\& \ge\!\overline{P}_{\rm ps,pd}+f P_{\rm ps,pd}\overline{P}_{\rm ps,ss} \min\bigg\{\frac{\lambda_{\rm e}}{\mu^{\left(o \right)}_{\rm e}},1\bigg\}\!=\!\mu^{\left(i \right)}_{\rm p}\!
 \end{split}
 \end{equation}

 The energy queue mean service rate provided in (\ref{mueexp}) can be upper bounded as follows. Since
  \begin{equation}
  \begin{split}
  &{\rm Pr}\bigg\{Q_{\rm p}\!=\!0,Q_{\rm s}\!=\!0,Q_{\rm ps}\!\ne\!0\bigg\}\!+\!{\rm Pr}\bigg\{Q_{\rm p}\!=\!0,Q_{\rm s}\!\ne\!0,Q_{\rm ps}\!=\!0\bigg\}\!\\& \,\,\,\,\,\,\,\,\,\,\,\,\,\,\,\,\,\,\,\,\ +\!{\rm Pr}\bigg\{Q_{\rm p}\!=\!0,Q_{\rm ps}\!\ne\!0,Q_{\rm ps}\!\ne\!0\bigg\}\le{\rm Pr}\bigg\{Q_{\rm p}=0\bigg\},
  \end{split}
  \end{equation}
      \begin{equation}
    \begin{split}
  \label{meuouter}
  \mu_{\rm e} &\le f{\rm Pr}\bigg\{Q_{\rm p}\ne0\bigg\}+{\rm Pr}\bigg\{Q_{\rm p}=0\bigg\} \\& \le 1-(1-f){\rm Pr}\bigg\{Q_{\rm p}\ne0\bigg\}\!\le\! 1\!-\!(1\!-\!f) \frac{\lambda_{\rm p}}{\mu_{\rm p}^{\left(o\right)}}=\mu_{\rm e}^{\left(o \right)}
  \end{split}
  \end{equation}
with ${\lambda_{\rm p}}\le{\mu^{\left(o \right)}_{\rm p}}$.

    The secondary source data queues ($Q_{\rm s}$ and $Q_{\rm ps}$) mean service rates are lower bounded as follows:
   \begin{equation}
   \begin{split}
   \label{88}
  \mu_{\rm ps}&=  \overline{P}_{\rm ss,pd}\biggr({\rm Pr}\bigg\{Q_{\rm p}=0,Q_{\rm s}=0,Q_{\rm e}\ne0\bigg\}\\& \,\,\,\,\,\,\,\,\,\,\,\,\,\,\,\,\,\,\,\,\ +\overline{\beta}{\rm Pr}\bigg\{Q_{\rm p}=0,Q_{\rm s}\ne0,Q_{\rm e}\ne0\bigg\}\biggr) \\&\ge^{E_1} \overline{P}_{\rm ss,pd}\biggr(\overline{\beta}{\rm Pr}\bigg\{Q_{\rm p}=0,Q_{\rm e}\ne0\bigg\}\biggr) \\& \ge  \overline{P}_{\rm ss,pd}\overline{\beta} (1-\frac{\lambda_{\rm p}}{\mu^{\left(i \right)}_{\rm p}})\min\bigg\{\frac{\lambda_{\rm e}}{\mu_{\rm e}^{\left(o \right)}},1\bigg\}=\mu_{\rm ps}^{\left(i \right)}
  \end{split}
  \end{equation}

    \begin{equation}
  \begin{split}
  \label{99}
  \mu_{\rm s} &= \overline{P}_{\rm ss,sd}\biggr({\rm Pr}\bigg\{Q_{\rm p}=0,Q_{\rm ps}=0,Q_{\rm e}\ne0\bigg\}\\& \,\,\,\,\,\,\,\,\,\,\,\,\,\,\,\,\,\,\,\,\,\,\,\,\,\,\,\,\,\,\,\ +\beta{\rm Pr}\bigg\{Q_{\rm p}=0,Q_{\rm ps}\ne0,Q_{\rm e}\ne0\bigg\}\biggr)\\&\ge^{E_2} \overline{P}_{\rm ss,sd}\biggr(\beta(1-\frac{\lambda_{\rm p}}{\mu^{\left(i \right)}_{\rm p}})\min\bigg\{\frac{\lambda_{\rm e}}{\mu_{\rm e}^{\left(o \right)}},1\bigg\}\biggr)=\mu_{\rm s}^{\left(i \right)}
  \end{split}
  \end{equation}
  with ${\lambda_{\rm p}}\le{\mu^{\left(i \right)}_{\rm p}}$. The inequalities $E_1$ and $E_2$ hold to equality under the assumption that $Q_{\rm s}$ and $Q_{\rm ps}$ never being empty, i.e., send dummy packets when they are empty. This assumption is similar to the dominant system. The stability of the dominant has been proved to be sufficient for the stability of the original system \cite{sadek,erph}. In our system, this assumption reduces the data queues service rates, since $Q_{\rm s}$ and $Q_{\rm ps}$ send dummy packets, therefore, when both queues are empty, there is a physical energy packet consumption per free time slot when $Q_{\rm p}=0$, and hence, the energy queue will be emptied faster without any contribution to the mean service rates of the queues. Moreover, the secondary source will not be able to help the primary source either by sending the future incoming relaying packet or by decoding more primary packets. Under this assumption, the first and the second inequalities of the mean service rate of the energy queue presented in (\ref{meuouter}) holds to equality. Therefore, sending dummy packets decreases the service rates of the data queues and increases the service rate of the energy queue.

  The mean arrival rate of the relaying queue is upper bounded as follows:
       \begin{equation}
     \begin{split}
   \lambda_{\rm ps}&=f  P_{\rm ps,pd}\overline{P}_{\rm ps,ss}{\rm Pr}\bigg\{Q_{\rm p}\ne0,Q_{\rm e}\ne0\bigg\}\\&  \le f P_{\rm ps,pd}\overline{P}_{\rm ps,ss}\min\bigg\{\frac{\lambda_{\rm e}}{\mu_{\rm e}^{\left(i \right)}},1\bigg\}\frac{\lambda_{\rm p}}{\mu^{\left(i \right)}_{\rm p}}=  \lambda_{\rm ps}^{\left(o \right)}
   \end{split}
   \end{equation}
   with ${\lambda_{\rm p}}\le{\mu^{\left(i \right)}_{\rm p}}$.
    \begin{table*}
\begin{center}
      \caption{The inner and the outer bounds of the mean arrival and service rates of the queues.}
\label{table1}
\begin{tabular}{ |c|c|c| }
    \hline
      $\mu_{\rm s}^{\left(i \right)}\!=\!\overline{P}_{\rm ss,sd}\beta(1\!-\!\frac{\lambda_{\rm p}}{\mu^{\left(i \right)}_{\rm p}})\min\bigg\{\frac{\lambda_{\rm e}}{\mu_{\rm e}^{\left(o \right)}},1\bigg\}$ & $\mu_{\rm s}^{\left(o \right)}\!=\!\overline{P}_{\rm ss,sd}(1\!-\!\frac{\lambda_{\rm p}}{\mu^{\left(o \right)}_{\rm p}}) \min\bigg\{\frac{\lambda_{\rm e}}{\mu_{\rm e}^{\left(i \right)}},1\bigg\}$ \\[3pt]\hline
      $\mu_{\rm e}^{\left(i \right)}\!=\!f \frac{\lambda_{\rm p}}{\mu_{\rm p}^{\left(o \right)}}$& $\mu_{\rm e}^{\left(o \right)}=1\!-\!(1\!-\!f) \frac{\lambda_{\rm p}}{\mu_{\rm p}^{\left(o\right)}}$ \\[3pt]\hline
      $\mu^{\left(i \right)}_{\rm p}\!=\!\overline{P}_{\rm ps,pd}+f P_{\rm ps,pd}\overline{P}_{\rm ps,ss} \min\bigg\{\frac{\lambda_{\rm e}}{\mu^{\left(o \right)}_{\rm e}},1\bigg\}$& $\mu_{\rm p}^{\left(o \right)}\!=\!\overline{P}_{\rm ps,pd}+f P_{\rm ps,pd}\overline{P}_{\rm ps,ss}$ \\[3pt]\hline $\mu_{\rm ps}^{\left(i \right)}=\overline{P}_{\rm ss,pd}\overline{\beta} (1-\frac{\lambda_{\rm p}}{\mu^{\left(i \right)}_{\rm p}})\min\bigg\{\frac{\lambda_{\rm e}}{\mu_{\rm e}^{\left(o \right)}},1\bigg\}$ & $\mu_{\rm ps}^{\left(o \right)}\!=\!\overline{P}_{\rm ss,pd}(1-\frac{\lambda_{\rm p}}{\mu^{\left(o \right)}_{\rm p}})\min\bigg\{\frac{\lambda_{\rm e}}{\mu_{\rm e}^{\left(i \right)}},1\bigg\}$ \\[3pt]\hline
      $ \lambda_{\rm ps}^{\left(i \right)}\!=\!f  P_{\rm ps,pd}\overline{P}_{\rm ps,ss}\min\bigg\{\frac{\lambda_{\rm e}}{\mu_{\rm e}^{\left(o \right)}},1\bigg\}\frac{\lambda_{\rm p}}{\mu^{\left(o \right)}_{\rm p}}$ & $  \lambda_{\rm ps}^{\left(o \right)}\!=\!f P_{\rm ps,pd}\overline{P}_{\rm ps,ss}\min\bigg\{\frac{\lambda_{\rm e}}{\mu_{\rm e}^{\left(i \right)}},1\bigg\}\frac{\lambda_{\rm p}}{\mu^{\left(i \right)}_{\rm p}} $ \\[3pt]\hline
\end{tabular}
\end{center}
\end{table*}

The outer and the inner bounds of the queues mean service and arrival rates are summarized in Table \ref{table1}. The outer and the inner bounds on the rates using ${\rm M/D/1}$ with unity service rate model for the energy queue can be obtained by following the same steps with $\mu_{\rm e}\!=\!\mu^{\left(i \right)}_{\rm e}\!=\!\mu^{\left(o \right)}_{\rm e}\!=\!1$ energy packets/slot.
  \subsection{Outer bound on the stability region}
 In order to obtain an outer bound on the stability region of the original system, we need to upper bound the service rates of the data queues and lower bound the service and arrival rates of the energy queue and relaying queue, respectively.
The outer bound's envelope points of the stability region is defined as the closure of the arrival rate pairs $(\lambda_{\rm p},\mu_{\rm s}^{\left(o \right)})$. More specifically, we fix $\lambda_{\rm p}$ and maximize
$\mu_{\rm s}^{\left(o \right)}$ as $f$ and $\beta$ vary over $[0,1]$ and under stability of all the queues in the network. The outer bound on the stable-throughput region of the system can be obtained by stating a constrained optimization problem as in \cite{sadek,erph}. Since the upper bounded mean service rates, $\mu^{\left(o \right)}_{\rm p}$ and $\mu^{\left(o \right)}_{\rm ps}$, and the lower bounded mean arrival rate to the relaying queue, $\lambda^{\left(i \right)}_{\rm ps}$, are independent of $\beta$ as shown in Table \ref{table1}, the optimization problem is expressed as follows:
  \begin{eqnarray}
  \begin{split}
  \label{opt1}
    \underset{0\le f\le 1}{\max.} &\ \ \ \ \lambda^{\left(o \right)}_{\rm s}=\mu_{\rm s}^{\left(o \right)},\ {\rm s.t.}  \ \ \ \    \lambda_{\rm p} \le \mu^{\left(o \right)}_{\rm p},\ \lambda^{\left(i \right)}_{\rm ps}\le\mu^{\left(o \right)}_{\rm ps}
    \end{split}
\end{eqnarray}
The upper bound optimization problem has only one optimization variable, $f$. The constraint $ \lambda_{\rm p} \le \mu^{\left(o \right)}_{\rm p}$ provides the lower value of $f$. That is, $\lambda_{\rm p}\le\overline{P}_{\rm ps,pd}+f P_{\rm ps,pd}\overline{P}_{\rm ps,ss}$. Hence, $f\ge \max\Biggr\{\frac{\lambda_{\rm p}- \overline{P}_{\rm ps,pd}}{P_{\rm ps,pd}\overline{P}_{\rm ps,ss}},0\Biggr\}$. If $\frac{\lambda_{\rm p}- \overline{P}_{\rm ps,pd}}{P_{\rm ps,pd}\overline{P}_{\rm ps,ss}}>1$, the optimization problem becomes infeasible. The optimal $f$ can be obtained via a simple grid search.

\subsection{Inner bound on the stability region}

In order to obtain an inner bound for the stability region, we need to lower the values of the service rates of the data queues, and elevate (increase) the values of the service rate of the energy queue and the arrival rate of the relaying queue. The inner bound on the stability region is obtained via solving the following optimization problem:
\begin{eqnarray}
\label{opt2}
\begin{split}
    \underset{0\le f,\beta\le 1}{\max.} & \ \ \ \ \lambda^{\left(i \right)}_{\rm s}=\mu_{\rm s}^{\left(i \right)},\ {\rm s.t.}   \ \ \ \   \lambda_{\rm p} \le \mu^{\left(i \right)}_{\rm p},\ \lambda^{\left(o \right)}_{\rm ps}\le\mu^{\left(i \right)}_{\rm ps}
    \end{split}
\end{eqnarray}
Fixing $f$, the optimization problem (\ref{opt2}) becomes a {\bf linear} program, i.e., a convex program. More specifically, for a given $f$, the objective function becomes linear in $\beta$, the primary queue constraint, $ \lambda_{\rm p} \le \mu^{\left(i \right)}_{\rm p}$, becomes constant in $\beta$, and the relaying queue stability constraint becomes linear in $\beta$. Hence, the optimization problem becomes linear in terms of $\beta$.

Rearranging the relaying queue stability constraint, $\beta$ is upper bounded as follows:
\begin{eqnarray}
\begin{split}
\beta \le 1-\frac{f P_{\rm ps,pd}\overline{P}_{\rm ps,ss}\min\bigg\{\frac{\lambda_{\rm e}}{\mu_{\rm e}^{\left(i \right)}},1\bigg\}\frac{\lambda_{\rm p}}{\mu^{\left(i \right)}_{\rm p}}}{ \overline{P}_{\rm ss,pd} (1-\frac{\lambda_{\rm p}}{\mu^{\left(i \right)}_{\rm p}})\min\bigg\{\frac{\lambda_{\rm e}}{\mu_{\rm e}^{\left(o \right)}},1\bigg\}}
    \end{split}
\end{eqnarray}
The optimization problem for a fixed $f$ is then given by
\begin{eqnarray}
\label{opt2xx}
\begin{split}
   & \underset{0\le \beta\le 1}{\max.}  \ \ \ \ \beta,\\& {\rm s.t.}   \ \ \ \  0\le \beta \le 1-\frac{f P_{\rm ps,pd}\overline{P}_{\rm ps,ss}\min\bigg\{\frac{\lambda_{\rm e}}{\mu_{\rm e}^{\left(i \right)}},1\bigg\}\frac{\lambda_{\rm p}}{\mu^{\left(i \right)}_{\rm p}}}{ \overline{P}_{\rm ss,pd} (1-\frac{\lambda_{\rm p}}{\mu^{\left(i \right)}_{\rm p}})\min\bigg\{\frac{\lambda_{\rm e}}{\mu_{\rm e}^{\left(o \right)}},1\bigg\}}
    \end{split}
\end{eqnarray}
 with $ \lambda_{\rm p}  \le \mu^{\left(i \right)}_{\rm p}$. The optimal $\beta$ is the highest feasible $\beta$. Hence, for a fixed $f$, the optimal $\beta$ is given by
\begin{eqnarray}
\begin{split}
\beta^*\!=\! 1-\frac{f P_{\rm ps,pd}\overline{P}_{\rm ps,ss}\min\bigg\{\frac{\lambda_{\rm e}}{\mu_{\rm e}^{\left(i \right)}},1\bigg\}\frac{\lambda_{\rm p}}{\mu^{\left(i \right)}_{\rm p}}}{ \overline{P}_{\rm ss,pd} (1-\frac{\lambda_{\rm p}}{\mu^{\left(i \right)}_{\rm p}})\min\bigg\{\frac{\lambda_{\rm e}}{\mu_{\rm e}^{\left(o \right)}},1\bigg\}}
    \end{split}
\end{eqnarray}
with $ \lambda_{\rm p} \le \mu^{\left(i \right)}_{\rm p}$. If $\beta^*<0$ and/or $ \lambda_{\rm p} > \mu^{\left(i \right)}_{\rm p}$, the problem is infeasible. If the optimization problem is infeasible, the secondary source has no access to the channel. The main idea of this approach is to solve a family of linear optimization problems parameterized by $f$. The optimal $f$ is chosen to be the one which achieves the highest objective function in (\ref{opt2}).
\section{Numerical Results}\label{numerical}
In this section, we provide some numerical results for the optimization problems presented in this paper. Let $\mathcal{S}^{(h)}$ and $\hat{\mathcal{S}}^{(h)}$ denote the proposed system when the secondary energy queue is modeled as a coupled queue with a certain service rate that depends on the other queues or as a decoupled queue with unity service rate, respectively. The superscript `$h$' reads `$\circ$' for outer bound and `$i$' for inner bound.

\begin{figure}
 \centering
  \includegraphics[width=1\columnwidth]{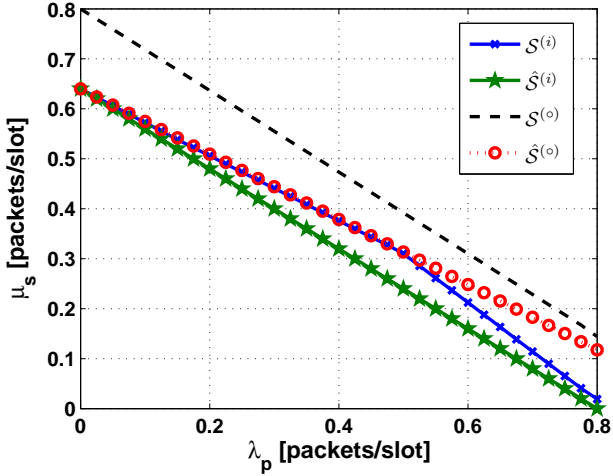}\\
  \caption{The stability region bounds of the proposed system. The parameters used to generate the figure are: $P_{\rm ps,pd}=0.2$,
$P_{\rm ss,sd}=0.2$, $P_{\rm ps,ss}=0.1$, $P_{\rm ss,pd}=0.1$ and $\lambda_{\rm e}=0.8$ energy packets per slot.}\label{fig1}
\end{figure}

In Fig. \ref{fig1}, we present a comparison between the bounds of the stability region of the proposed cooperative system under the proposed energy queue model and the proposed cooperative system in which the energy queue is modeled as a decoupled ${\rm M/D/1}$ with unity service rate. In the decoupled ${\rm M/D/1}$ with unity service rate model, an energy packet is consumed every time slot. In particular, the mean service rate of the energy queue is $\mu_{\rm e}=1$ energy packets/slot regardless of the secondary queues state and the secondary operations; hence, it is a non-realistic assumption as mentioned earlier. The parameters used to generate the figure are: $P_{\rm ps,pd}=0.2$,
$P_{\rm ss,sd}=0.2$, $P_{\rm ps,ss}=0.1$, $P_{\rm ss,pd}=0.1$ and $\lambda_{\rm e}=0.8$ energy packets/slot.

\begin{figure}
  \centering
  \includegraphics[width=1\columnwidth]{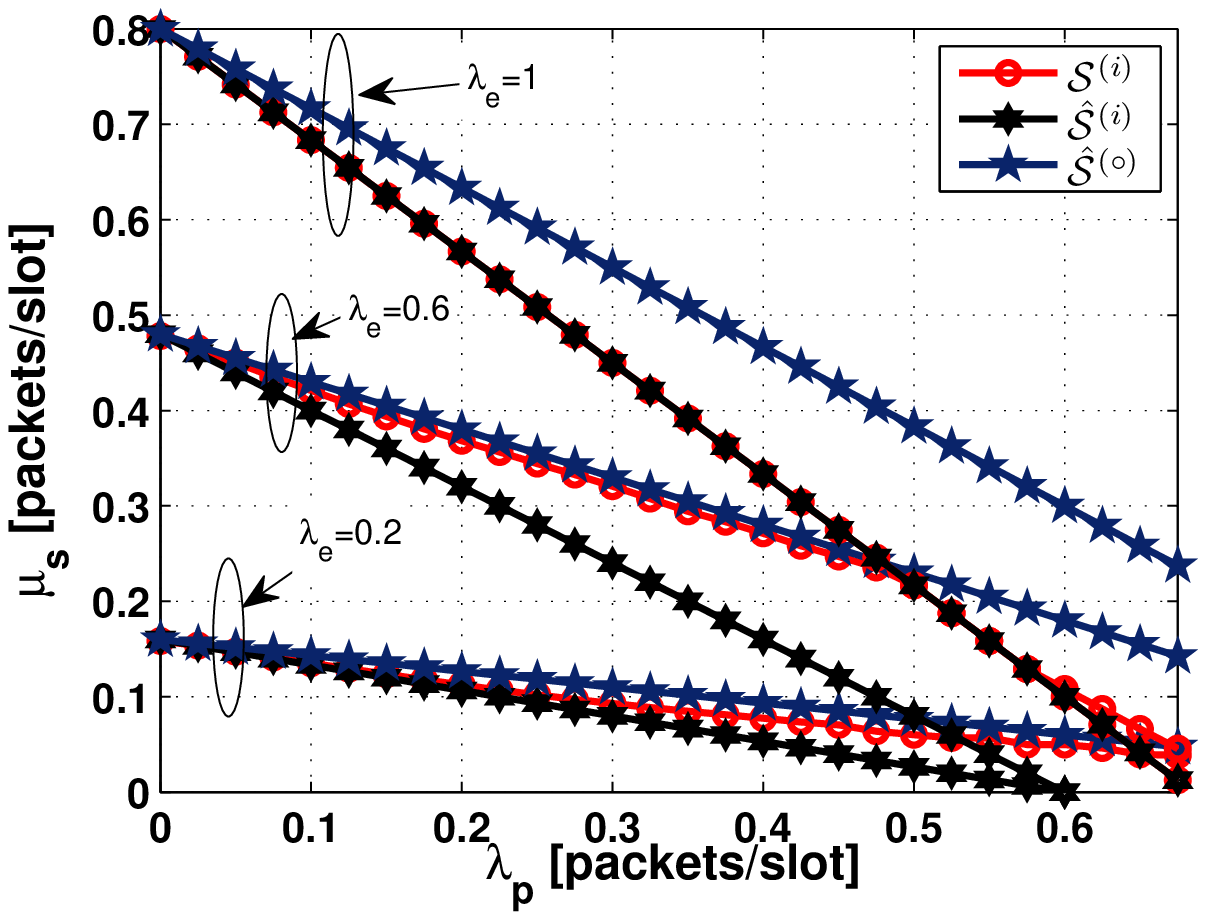}\\
  \caption{The stability region bounds of the proposed system. The parameters used to generate the figure are: $P_{\rm ps,pd}=0.4$,
$P_{\rm ss,sd}=0.2$, $P_{\rm ps,ss}=0.1$, and $P_{\rm ss,pd}=0.1$.}\label{fig2}
\end{figure}

Fig. \ref{fig2} shows the impact of the arrival rate of the energy queue on the stability bounds. More specifically, the figure compares the inner bound of the proposed system with and without modeling the energy queue as a decoupled ${\rm M/D/1}$ queue with unity service rate. The parameters used to generate the figure are: $P_{\rm ps,pd}=0.4$,
$P_{\rm ss,sd}=0.2$, $P_{\rm ps,ss}=0.1$, and $P_{\rm ss,pd}=0.1$.

 From Figs. \ref{fig1} and \ref{fig2}, it is noted that $\mathcal{S}^{(i)}$ almost overlaps with $\hat{\mathcal{S}}^{(\circ)}$, specifically, for $\lambda_{\rm e}<1$ energy packets/slot. This is because the decoupled ${\rm M/D/1}$ with unity service rate model assumes an energy packet consumption per slot even if the secondary source is silent (in terms of decoding a primary packet or accessing a sensed free time slot with a data packet) and therefore, under this assumption, the energy queue will be emptied rapidly and the secondary source will not be able to accept packets from PU or to deliver its own data packets. For $\lambda_{\rm e}=1$ energy packets/slot, the bounds of the proposed model and the ${\rm M/D/1}$ with unity service rate model coincide because the probability of the energy queue being nonempty for both models is $1$ (there is an energy arrival every slot and therefore the energy queue is always full). This case presents an SU connected to a reliable power supply.

\begin{figure}
  \centering
  \includegraphics[width=1\columnwidth]{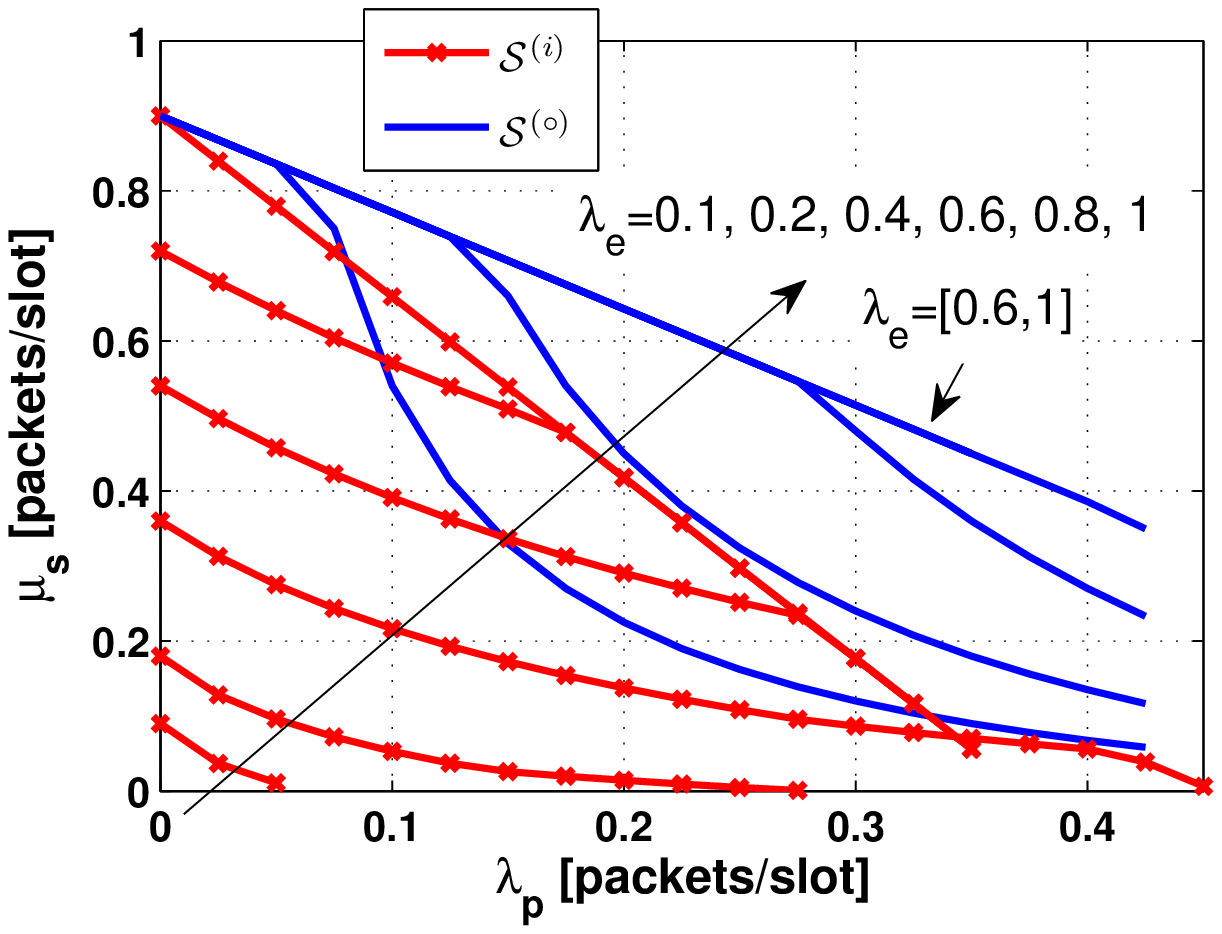}\\
  \caption{The stability region bounds of the proposed system. The parameters used to generate the figure are: $P_{\rm ps,pd}=1$,
$P_{\rm ss,sd}=0.1$, $P_{\rm ps,ss}=0.3$, and $P_{\rm ss,pd}=0.2$.}\label{fig3}
\end{figure}

Fig. \ref{fig3} shows the impact of the arrival rate of the energy queue on the proposed system bounds. The parameters used to generate the figure are: $P_{\rm ps,pd}=1$,
$P_{\rm ss,sd}=0.1$, $P_{\rm ps,ss}=0.3$, $P_{\rm ss,pd}=0.2$, and the values of $\lambda_{\rm e}$ shown in the figure in energy packets/slot. For both bounds, we plot the figure for six values of $\lambda_{\rm e}$ energy packets/slot. The outer bounds of both models are overlapped for $\lambda_{\rm e}\in[0.6,1]$. Note that without cooperation the SU will not gain free slots to access the channel and the primary queue will be unstable (never being empty). This is because the mean service rate of the primary queue, in case of noncooperative SU, is $1-P_{\rm ps,pd}=0$. The stability region bounds expand with increasing the secondary energy arrival rate. That is, the feasible range of the primary mean arrival rate and the maximum secondary throughput increase for each feasible $\lambda_{\rm p}$. This is because there will be sufficient energy for both helping the PU and transmitting the packets from both the relaying and data queues.

\begin{figure}
  \centering
  \includegraphics[width=1\columnwidth]{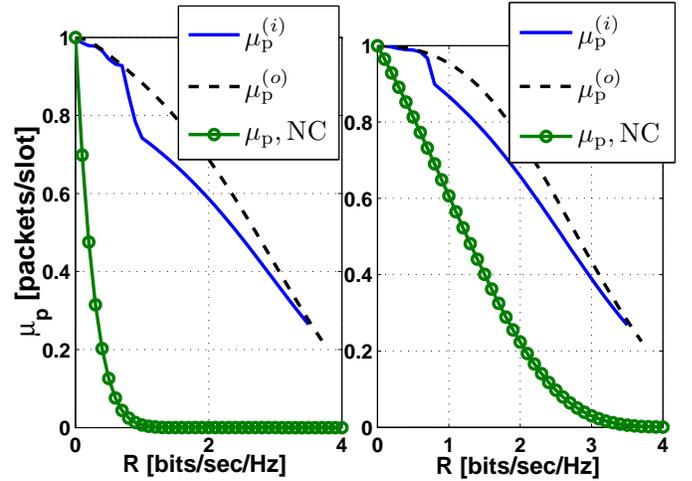}\\
  \caption{The maximum primary service rate versus spectral efficiency $R=b/(TW)$ in bits per second per hertz (bits/s/Hz). The left subfigure is plotted with $\gamma_{\rm ps,pd}=0.2$, whereas the right subfigure is plotted with $\gamma_{\rm ps,pd}=2$. The non-cooperative system is denoted by `NC'.}\label{fig4}
\end{figure}

In Fig. \ref{fig4}, the maximum primary service rate is plotted against the spectral efficiency
$R=\frac{b}{TW}$. The inner and the outer bounds on the primary mean service rate of the proposed system are compared with the non-cooperation case. The advantage of the cooperative system over the non-cooperative (NC) system is noted. For this
case, the proposed system's bounds start from $1$ for low rates $R$ but decays
exponentially after that. The parameters used to generate the figure are: $\gamma_{\rm ps,ss}=8$, $\gamma_{\rm ss,sd}=8$, $\gamma_{\rm ss,sd}=8$, $\tau=0.1 T$ seconds, $\lambda_{\rm e}=0.7$ energy packets per slot, and $\lambda_{\rm p}=0.1$ packets per slot.

\section{Conclusions}\label{conc}

In this paper, we have investigated the impact of having an energy harvesting cognitive radio user with a PU. We have provided an inner bound and an outer bound on the stability region of the proposed cooperative system. In contrast to the decoupled ${\rm M/D/1}$ energy queue with unity service rate model \cite{pappas2012optimal}, we have modeled the energy queue as a discrete time queue with certain arrival rate and certain service rate. The service process depends on the the behavior of the SU and its data queue states. The beneficial gains of cooperation are noted for both the primary and the secondary data queues in terms of mean service rates. We have investigated the impact of the energy queue arrival rate on the secondary and primary mean service rates. A possible extension of this work is to consider a multipacket reception channel model where concurrent transmissions are allowed. In this case, the SU can transmit concurrently with the PU to exploit the multipacket reception capabilities of receivers.
\section*{Acknowledgement}
This research work is funded by Qatar National Research Fund (QNRF) under grant number NPRP 6-1326-2-532.
\bibliographystyle{IEEEtran}
\bibliography{IEEEabrv,wcm_bib}

\begin{thebibliography}{10}
\providecommand{\url}[1]{#1}
\csname url@samestyle\endcsname
\providecommand{\newblock}{\relax}
\providecommand{\bibinfo}[2]{#2}
\providecommand{\BIBentrySTDinterwordspacing}{\spaceskip=0pt\relax}
\providecommand{\BIBentryALTinterwordstretchfactor}{4}
\providecommand{\BIBentryALTinterwordspacing}{\spaceskip=\fontdimen2\font plus
\BIBentryALTinterwordstretchfactor\fontdimen3\font minus
  \fontdimen4\font\relax}
\providecommand{\BIBforeignlanguage}[2]{{%
\expandafter\ifx\csname l@#1\endcsname\relax
\typeout{** WARNING: IEEEtran.bst: No hyphenation pattern has been}%
\typeout{** loaded for the language `#1'. Using the pattern for}%
\typeout{** the default language instead.}%
\else
\language=\csname l@#1\endcsname
\fi
#2}}
\providecommand{\BIBdecl}{\relax}
\BIBdecl

\bibitem{haykin2005cognitive}
S.~Haykin, ``{Cognitive} radio: brain-empowered wireless communications,''
  \emph{IEEE J. Sel. Areas Commun.}, vol.~23, no.~2, pp. 201--220, Feb. 2005.

\bibitem{ref1}
L.~Xiang, X.~Ge, C.-X. Wang, F.~Y. Li, and F.~Reichert, ``Energy efficiency
  evaluation of cellular networks based on spatial distributions of traffic
  load and power consumption,'' \emph{IEEE Trans. Wireless Commun.}, vol.~12,
  no.~3, pp. 961--973, March 2013.

\bibitem{ref2}
I.~Humar, X.~Ge, L.~Xiang, M.~Jo, M.~Chen, and J.~Zhang, ``Rethinking energy
  efficiency models of cellular networks with embodied energy,'' \emph{IEEE
  Network}, vol.~25, no.~2, pp. 40--49, March 2011.

\bibitem{ref3}
X.~Ge, J.~Hu, C.-X. Wang, C.-H. Youn, J.~Zhang, and X.~Yang, ``Energy
  efficiency analysis of {MISO-OFDM} communication systems considering power
  and capacity constraints,'' \emph{Mobile Networks and Applications}, vol.~17,
  no.~1, pp. 29--35, 2012.

\bibitem{c0}
Z.~Ding, I.~Krikidis, B.~Sharif, and H.~V. Poor, ``Wireless information and
  power transfer in cooperative networks with spatially random relays.''\hskip
  1em plus 0.5em minus 0.4em\relax Available
  [Online]:http://arxiv.org/abs/1403.6164.pdf, 2014.

\bibitem{c1}
H.~Chen, Y.~Li, J.~L. Rebelatto, B.~F. Uchoa-Filhoand, and B.~Vucetic,
  ``Harvest-then-cooperate: Wireless-powered cooperative
  communications.''\hskip 1em plus 0.5em minus 0.4em\relax Available
  [Online]:http://arxiv.org/abs/1404.4120.pdf, 2014.

\bibitem{survey}
S.~Sudevalayam and P.~Kulkarni, ``Energy harvesting sensor nodes: Survey and
  implications,'' \emph{IEEE Communications Surveys and Tutorials}, vol.~13,
  no.~3, pp. 443--461, 2011.

\bibitem{hoang2009opportunistic}
A.~Hoang, Y.~Liang, D.~Wong, Y.~Zeng, and R.~Zhang, ``Opportunistic spectrum
  access for energy-constrained cognitive radios,'' \emph{IEEE Trans. Wireless
  Commun.}, vol.~8, no.~3, pp. 1206--1211, Mar. 2009.

\bibitem{sharma2010optimal}
V.~Sharma, U.~Mukherji, V.~Joseph, and S.~Gupta, ``Optimal energy management
  policies for energy harvesting sensor nodes,'' \emph{IEEE Trans. Wireless
  Commun.}, vol.~9, no.~4, pp. 1326--1336, Apr. 2010.

\bibitem{ho2010optimal}
C.~Ho and R.~Zhang, ``Optimal energy allocation for wireless communications
  powered by energy harvesters,'' in \emph{IEEE ISIT}, Jun. 2010, pp.
  2368--2372.

\bibitem{yang2010transmission}
J.~Yang and S.~Ulukus, ``Transmission completion time minimization in an energy
  harvesting system,'' in \emph{IEEE 44th CISS}, Mar., pp. 1--6.

\bibitem{yang2010optimal}
------, ``Optimal packet scheduling in an energy harvesting communication
  system,'' \emph{IEEE Trans. on Commun.}, vol.~60, no.~1, pp. 220--230, Jan.
  2012.

\bibitem{tutuncuoglu2010optimum}
K.~Tutuncuoglu and A.~Yener, ``Optimum transmission policies for battery
  limited energy harvesting nodes,'' \emph{IEEE Trans. Wireless Commun.},
  vol.~11, no.~3, pp. 1180--1189, Mar. 2012.

\bibitem{pappas2012optimal}
N.~Pappas, J.~Jeon, A.~Ephremides, and A.~Traganitis, ``Optimal utilization of
  a cognitive shared channel with a rechargeable primary source node,'' in
  \emph{JCN}, vol.~14, no.~2, Apr. 2012, pp. 162--168.

\bibitem{krikidis2012stability}
I.~Krikidis, T.~Charalambous, and J.~Thompson, ``Stability analysis and power
  optimization for energy harvesting cooperative networks,'' \emph{IEEE Signal
  Process. Lett.}, vol.~19, no.~1, pp. 20--23, Jan. 2012.

\bibitem{Sult1210:Optimal}
A.~{El Shafie} and A.~Sultan, ``Optimal random access and random spectrum
  sensing for an energy harvesting cognitive radio,'' in \emph{Proc. WiMob},
  Oct. 2012, pp. 403--410.

\bibitem{Sultan}
A.~Sultan, ``Sensing and transmit energy optimization for an energy harvesting
  cognitive radio,'' \emph{IEEE Wirel. Commun. Lett.}, vol.~1, no.~5, pp.
  500--503, Oct. 2012.

\bibitem{gc2013}
A.~{El Shafie} and A.~Sultan, ``Optimal selection of spectrum sensing duration
  for an energy harvesting cognitive radio,'' in \emph{Proc. IEEE GLOBECOM},
  2013, pp. 1020--1025.

\bibitem{ref4}
A.~Goldsmith, S.~A. Jafar, I.~Maric, and S.~Srinivasa, ``Breaking spectrum
  gridlock with cognitive radios: An information theoretic perspective,''
  \emph{Proceedings of the IEEE}, vol.~97, no.~5, pp. 894--914, 2009.

\bibitem{ref5}
Y.~Han, A.~Pandharipande, and S.~H. Ting, ``Cooperative decode-and-forward
  relaying for secondary spectrum access,'' \emph{IEEE Trans. Wireless
  Commun.}, vol.~8, no.~10, pp. 4945--4950, 2009.

\bibitem{ref6}
Q.~Li, S.~H. Ting, A.~Pandharipande, and Y.~Han, ``Cognitive spectrum sharing
  with two-way relaying systems,'' \emph{IEEE Trans. Veh. Technol.}, vol.~60,
  no.~3, pp. 1233--1240, 2011.

\bibitem{ref7}
Q.~Li, S.~H. Ting, A.~Pandharipande, and M.~Motani, ``Cooperate-and-access
  spectrum sharing with {ARQ}-based primary systems,'' \emph{IEEE Trans.
  Commun.}, vol.~60, no.~10, pp. 2861--2871, 2012.

\bibitem{simeone}
O.~Simeone, Y.~Bar-Ness, and U.~Spagnolini, ``Stable throughput of cognitive
  radios with and without relaying capability,'' \emph{IEEE Trans. on Commu.},
  vol.~55, no.~12, pp. 2351--2360, Dec. 2007.

\bibitem{khattab}
M.~Elsaadany, M.~Abdallah, T.~Khattab, M.~Khairy, and M.~Hasna, ``Cognitive
  relaying in wireless sensor networks: Performance analysis and
  optimization,'' in \emph{Proc. IEEE GLOBECOM}, Dec. 2010, pp. 1--6.

\bibitem{erph}
S.~Kompella, G.~Nguyen, J.~Wieselthier, and A.~Ephremides, ``Stable throughput
  tradeoffs in cognitive shared channels with cooperative relaying,'' in
  \emph{Proc. IEEE INFOCOM}, Apr. 2011, pp. 1961--1969.

\bibitem{sadek}
A.~Sadek, K.~Liu, and A.~Ephremides, ``Cognitive multiple access via
  cooperation: protocol design and performance analysis,'' \emph{IEEE Trans.
  Inf. Theory}, vol.~53, no.~10, pp. 3677--3696, Oct. 2007.

\bibitem{tsy}
B.~Tsybakov and W.~Mikhailov, ``Ergodicity of slotted {ALOHA} system,''
  \emph{Problemy Peredachi Informatsii}, vol.~15, no.~4, pp. 73--–87, 1979.

\bibitem{rao1988stability}
R.~Rao and A.~Ephremides, ``On the stability of interacting queues in a
  multiple-access system,'' \emph{IEEE Trans. Inf. Theory}, vol.~34, no.~5, pp.
  918--930, Sept. 1988.

\bibitem{szpankowski1994stability}
W.~Szpankowski, ``Stability conditions for some distributed systems: buffered
  random access systems,'' \emph{Advances in Applied Probability}, pp.
  498--515, 1994.

\bibitem{luo1999stability}
W.~Luo and A.~Ephremides, ``Stability of {N} interacting queues in
  random-access systems,'' \emph{IEEE Trans. Inf. Theory}, vol.~45, no.~5, pp.
  1579--1587, July 1999.

\bibitem{naware2005stability}
V.~Naware, G.~Mergen, and L.~Tong, ``Stability and delay of finite-user slotted
  {ALOHA} with multipacket reception,'' \emph{IEEE Trans. Inf. Theory},
  vol.~51, no.~7, pp. 2636--2656, July 2005.

\end{thebibliography}
\balance
\end{document}